\documentclass[aps,prb,amsmath,amssymb,reprint,showkeys]{revtex4-2}

\usepackage[dvipsnames]{xcolor}
\usepackage{graphicx}
\usepackage{hyperref}
\usepackage[utf8]{inputenc}
\usepackage[T1]{fontenc}
\usepackage{physics}
\usepackage{mathtools}
\usepackage{siunitx}
\usepackage{bm}
\usepackage{dcolumn}
\usepackage{appendix}
\usepackage{lipsum}
\usepackage{multirow}

\usepackage[normalem]{ulem}
\newcommand{\Review}[1]{\textcolor{black}{#1}}

\begin{document}
\title{A spectral-neighbour representation for vector fields: machine-learning potentials including spin}
\author{M.~Domina}
\author{M.~Cobelli}
\author{S.~Sanvito}
\affiliation{School of Physics and CRANN Institute, Trinity College Dublin, Ireland}
\date{\today}
\begin{abstract}
We introduce a translational and rotational invariant local representation for vector fields, which can be employed 
in the construction of machine-learning energy models of solids and molecules. This allows us to describe, on the 
same footing, the energy fluctuations due to the atomic motion, the longitudinal and transverse excitations of the 
vector field, and their mutual interplay. The formalism can then be applied to physical systems where the total 
energy is determined by a vector density, as in the case of magnetism. Our representation is constructed over the 
power spectrum of the combined angular momentum describing the local atomic positions and the vector field, and 
can be used in conjunction with different machine-learning schemes and data taken from accurate {\it ab initio} 
electronic structure theories. We demonstrate the descriptive power of our representation for a range of classical 
spin Hamiltonian and machine-learning algorithms. In particular, we construct energy models based on both linear 
Ridge regression, as in conventional spectral neighbour analysis potentials, and gaussian approximation. These 
are both built to represent a Heisenberg-type Hamiltonian including a longitudinal energy term and spin-lattice coupling.
\end{abstract}
\maketitle

\section{Introduction}

The modelling of the structural, electronic and magnetic properties of materials at finite temperature requires the 
exploration of complex energy surfaces, a task that is usually performed in the configuration space or through 
extensive time-dependent simulations. The gold standard is set by {\it ab initio} methods, where one solves directly 
an electronic problem, for instance through density functional theory. The accuracy of the method is then given by 
the accuracy of the underline electronic structure theory, which often can be determined by the level of approximation 
taken. Notably, also the computational overheads are set by the electronic structure theory, so that there is always a 
trade off between the accuracy of the prediction, the duration of the simulation and the maximum size of the system 
to simulate. Real-time time-dependent simulations of purely electronic quantities, such as the dynamics of 
spins~\cite{Elliott2016,Simoni2017,Simoni2017a}, are limited to a few atoms and a few hundreds of femtoseconds, 
while {\it ab initio} molecular dynamics (MD) simulations can reach well within the picosecond range and may 
involve several hundreds of atoms.

The general strategy for extending the range of dynamics simulations across both the time and length scale is to 
abandon completely the {\it ab initio} description and replace the solution of the electronic problem with some 
parametric functions, constructed to reproduce the {\it ab-initio} potential energy surface, namely classical 
force fields~\cite{Frenkel2002}. In their most canonical form one describes the interaction among the ions by 
introducing energy contributions that account for the various physical forces at play (covalent bond, dispersive 
forces, etc.). Once the force field is defined, only the atomic positions determine the total energy. A similar approach
has been recently introduced for spin-dynamics. In this case one associates to each atom a classical spin vector,
$\mathbf{S}_i$, so that the total energy is defined over a continuous vector field with values at the atomic 
positions~\cite{Evans2014,Eriksson2017}. The total energy then takes the form of a classical Heisenberg model and 
may include both anisotropy and friction terms. Furthermore, the formalism can be extended to include spin-lattice 
coupling~\cite{Ma2018}, longitudinal spin fluctuations~\cite{PhysRevB.86.054416}, and possibly the effects of a 
spin current~\cite{Ellis2017}. 

In general, force fields constructed in this way have two main drawbacks. On the one hand, their accuracy is 
significantly inferior to that of an {\it ab initio} electronic structure theory, although this varies depending on the 
class of compounds one wants to study. On the other hand, they tend to be specific to the particular type of 
bond they describe. In addition, spin-type force fields may not be able to describe entire excitations types. For 
instance, magnetic Stoner excitations are not part of the spectrum of a classical Heisenberg model. 

Recently, a new class of force fields, named machine-learning force fields (MLFFs), have been shown to solve both 
the accuracy and specificity issues. The general philosophy of MLFFs is quite different from that of their classical 
counterparts, since one does not pretend to construct an energy function of the atomic coordinates with terms baring
a physical interpretation, but instead tries to reproduce extremely accurately the {\it ab initio} potential energy surface.
MLFFs comprise of two parts, an abstract representation of the atom density distribution~\cite{Musil2021} and a 
machine-learning model that correlates such representation to the system total energy. For the two parts to work together
the representation should be translational, atom-permutational and rotational invariant. The first two conditions are
usually met by local representations (also known as atomic-neighbour descriptions), where the energy is expressed in 
terms of atomic contributions, while the last condition is constructed in. Symmetry functions~\cite{Behler-Parrinello}
and bispectra~\cite{Bartok} are two examples of such rotational invariant local representations. 

\Review{Importantly, the spin degrees of freedom are not explicitly included in the representation, which typically describes 
an atomic density only. For this reason MLFFs are currently unable to describe the energy difference between inequivalent 
magnetic phases, for instance between a ferromagnetic and an antiferromagnetic ground state, unless the different phases 
are also associated to different structures. This limitation can be addressed by combining the MLFF with a classic spin 
Hamiltonian in order to create a model able to predict energies dependent on both the atomic positions and the spin order.
This route was recently explored by Nikolov and co-workers \cite{Nikolov2021}, who equipped a spectral neighbour analysis 
potential (SNAP) with a classical Heisenberg Hamiltonian to compute thermodynamical properties (e.g. the Curie
temperature) of $\alpha$-Fe.}

\Review{Another possible strategy is to make the MLFF aware of the spin configuration by including such information in 
the input features of the model. This is a non trivial task, since such features need to retain the aforementioned symmetries 
in order for the model to perform well. A recent attempt along this direction consists in the introduction of a novel definition
of symmetry functions carrying spin information~\cite{Eckhoff2021}. Such reformulation was constructed for the spin-collinear 
case and for fixed spin magnitude (no longitudinal spin information is available). Similarly, the atomic-cluster expansion method 
was recently extended to vector fields~\cite{vectorACE}, in a way that enables the description of non-collinear spin configurations. 
Notably, by design both methods require a large number of features for an accurate description of the magnetic environment, 
making them particularly data hungry. Thus, a compact representation describing a vector field, hence able to compute atomic 
and magnetic excitations on the same footing, still remains at large. Our paper aims to fill the gap.}

Here, we propose a new local representation for vector fields, which can be used with either linear and non-linear 
machine-learning models. This is based on the power spectrum of the combined angular momentum describing the 
local atomic positions and the vector field. The representation is rotationally invariant and can be further generalised
to tensorial densities. In order to test its descriptive power, such representation is combined with either linear Ridge 
regression and gaussian approximation to construct MLFFs describing the potential energy surface of a Heisenberg 
model with longitudinal fluctuations and spin-lattice coupling. Our results show that extremely accurate energy predictions 
can be obtained with a rather moderate number of training data. 

\section{Methods}

\subsection{Density for a vector field}
The starting point of any atomic-neighbour description of solids and molecules consists in defining the 
local particle density associated to the $i$-th atom,
\begin{equation}\label{local_rho}
\rho_i(\vb r) = \sum_{r_{ai} < r_\text{cut}} w_a h_{a}(\bm r-\bm r_{ai})\:.
\end{equation}
Here, $\bm r_{ai} = \bm r_a-\bm r_i$ is the distance between the atoms at the position $\bm r_a$ and
$\bm r_i$, $r_{ai}=|{\bm r_{ai}}|$, so that the coordinates of the $i$-th atom define the origin of the 
local reference frame. The sum in Eq.~(\ref{local_rho}) runs over all the atoms inside a sphere of 
radius $r_\text{cut}$ with center at $\bm r_i$, while $w_a$ are weights usually associated to the atomic 
species of the $a$-th atom. In this expansion $h_{a}$ is a localisation function, such a Gaussian or a 
Dirac-delta, centred at $\bm r_a$, whose specific shape, in general, can depend on the $a$-th atom type. 
Atomic-neighbour descriptions are then constructed by defining rotationally invariant combinations of the
coefficients of expansion of $h_{a}$ over an appropriate local basis \cite{Musil2021}.

In the same spirit, we can now define a local \emph{vector} density, $\bm \rho(\vb r)$, through 
Eq.~(\ref{local_rho}) by associating a vector $\bm v_{a}$ to each position $\bm r_a$, namely
\begin{equation}\label{eq:2.1}
\bm \rho(\vb r) = \sum_{a} w_a h_{a}(\bm r-\bm r_a)\bm v_{a}\:,
\end{equation}  
where for simplicity we have dropped the index $i$. In this formulation $\bm v_{a}$ may, for instance,
represent the local moment of the ions in a magnetic compound, so that $\bm \rho(\vb r)$ describes the
local magnetisation field. In Eq~(\ref{eq:2.1}) the vector $\bm v_{a}$ is defined through its cartesian
components
\begin{equation}\label{eq:2.2}
\bm v_a = v_{a,x} \hat{\bm e}_x+v_{a,y} \hat{\bm e}_y+ v_{a,z} \hat{\bm e}_z= \sum_{i =x,y,z} v_{a,i}\hat{\bm e}_i\:,
\end{equation}
with $\hat{\bm e}_i$ being the unit vector along $i=x, y, z$. However, it is convenient to replace the decomposition 
of Eq.~\eqref{eq:2.2} with one using the spherical versors \cite{Weissbluth} ,
\begin{equation}\label{eq:2.5}
\hat{\bm e}_{\pm 1} = \mp\dfrac{1}{\sqrt 2}(\hat{\bm e}_x \pm i\hat{\bm e}_y)\quad\text{and}\quad \hat{\bm e}_0 = \hat{\bm e}_z\:,
\end{equation}
so $\bm v_a$ becomes,
\begin{equation}\label{eq:2.6}
\bm v_a = \sum_{q = 0,\pm 1} v_{a,q} \hat{\bm e}_q\quad\text{with}\quad
\begin{dcases}
v_{a,\pm 1} = \mp\dfrac{1}{\sqrt 2}(v_{a,x}\mp iv_{a,y})\:,\\
v_{a,0} = v_{a,z}\:.
\end{dcases}
\end{equation}
This decomposition is a particular case of the more general one for a tensor of order $\lambda$ in its
irreducible spherical components. Therefore, the spherical components $v_{a,q}$ transforms under rotations 
as the spherical harmonic $Y^{q}_1$ \cite{Weissbluth,AngularMomentum}.

In order to construct covariant descriptors for the local vector density of Eq.~\eqref{eq:2.1}, one first needs
to expand the spatial part, $h_a$, over an orthonormal radial basis. Here we use the product between a radial
basis, $R_{nl}(r)$, and the three-dimensional spherical harmonics, $Y_l^m(\hat{\bm r})$, where as usual $n$, $l$ 
and $m$ are, respectively, the principal, the angular momentum and the third component of the angular momentum 
quantum numbers. The local vector density then becomes, 
\begin{equation}\label{eq:2.3}
\bm \rho(\bm r) \simeq \sum_{n=0}^{n_\text{max}}\sum_{l=0}^{n} \sum_{m=-l}^l \sum_{q=0,\pm 1}c_{nlmq}R_{nl}(r)Y_l^m (\hat{\bm r})\hat{\bm e}_q\:,
\end{equation} 
where the equality holds only for a complete basis but not for the one truncated at $n_\text{max}$, and where
the coefficients of expansion are calculated as
\begin{equation}\label{eq:2.4}
c_{nlmq} = \sum_a w_a v_{a,q}\int \dd r \dd \hat{\bm r} \: r^2 R_{nl}(r){Y_l^m }^*(\hat{\bm r})h_a(\bm r-\bm r_a)\:.
\end{equation}
In this work we choose the radial-basis set introduced for the Spherical-Bessel descriptors~\cite{SBdescriptors}, 
namely $R_{nl}(r) = g_{n-l,l}(r)$. These are orthonormal on the sphere and smoothly vanish at the cut-off radius. 
Note that the choice of the radial basis set is not unique or crucial and alternative basis can be selected. The only
practical criterion is that they should approximate completeness with a relative small number of basis functions, 
namely the convergence must be rapid. Note also that, in what follows, we will always assume orthonormality within 
the radial basis set, although the expressions derived could be easily generalised to the non-orthogonal case.

The use of the spherical components of a tensor has been already exploited in the construction of covariant 
kernels for vectorial and a tensorial properties related to an atomic 
environment~\cite{PhysRevLett.120.036002,PhysRevB.95.214302,Ale}. 
Here, we follow the same basic idea and formalism, which stems from the decomposition of a tensorial object into 
spherical components. The main difference is that we cannot just associate a tensorial object to our density, 
since the density itself is the vectorial field. As such, we will avail of the same concepts and methods that are 
typically used to describe the coupling of angular momenta and vector fields in atomic physics~\cite{Weissbluth}. 
Another difference with existing literature is that our primary target is the construction of invariant quantities 
instead of covariant ones. A similar goal has been already pursued in Ref.~\cite{vectorACE}, where a vector 
field was described using the same strategy employed in dealing with the atomic positions. More explicitly, 
in \cite{vectorACE} the magnetic vectors are encoded in Dirac's delta distributions, which are then expanded 
on a suitable basis set and coupled with the analogous expansion arising from the positions of the atoms. Here, 
however we will preserve the vectorial nature of the field at each step in the derivation, a strategy that results in 
a simpler coupling scheme, as it will be shown in detail below. With this in mind, we will first proceed with deriving 
an invariant power spectrum for the vector density of Eq.~\eqref{eq:2.1}, and then we will implement a linear regression 
to fit such power spectrum. This second step is similar to what is commonly done with the formalism of the 
spectral neighbour analysis potentials (SNAPs)~\cite{SNAP}.  

For the remaining of the paper we will develop the formalism by using a conventional Dirac notation, which 
allows one to appreciate better the structure of the representation. In fact, as shown in reference \cite{Ceriotti}, 
the Dirac notation gives us a natural tool for dealing with local atomic densities. In this way the expansion of 
the vector density defined in Eq.~\eqref{eq:2.3} can be written in a compact form as
\begin{equation}\label{eq:2.7}
\ket{\bm \rho} = \sum_{nlmq}c_{nlmq}\ket{nlmq}\:,
\end{equation}
where 
\begin{equation}\label{eq:2.8}
\braket{\bm r}{nlm} = g_{n-l,l}(r)Y^{m}_l(\hat{\bm r}),\quad\text{and}\quad \ket{q} \equiv \hat{\bm e}_q\:.
\end{equation}
We can then express all the relevant quantities over the the basis, $\ket{nl1JM}$, of the combined angular 
momenta, $\bm L+\bm 1= \bm J$, by using the standard addition scheme \cite{AngularMomentum},
\begin{equation}\label{eq:2.9}
\ket{nl1JM} = \sum_{m = -l}^l \sum_{q= -1}^1 C^{JM}_{lm1q} \ket{nlmq},
\end{equation}
where $C^{JM}_{lm1q}$ are the Clebsch-Gordan coefficients. As usual, $J$ and $M$, are the quantum numbers
for the total angular momentum and its projection, while `1' refers to the angular momentum of $\bm 1$, describing 
the vector nature of the field. Hence, we have
\begin{equation}\label{eq:2.10}
\abs{l-1} \leq J \leq{l+1}\quad\text{and}\quad -J \leq M \leq J\:.
\end{equation}
The states $\ket{nlJM}$, when projected over the position representation, are the products of \emph{vector} 
spherical harmonics and radial functions. By inverting equation \eqref{eq:2.9},
\begin{equation}\label{eq:2.11}
\ket{nlmq} = \sum_{JM} C^{JM}_{lm1q} \ket{nlJM}\:,
\end{equation}
we can write the vector density as
\begin{equation}\label{eq:2.12}
\ket{\bm \rho} = \sum_{nlJM} u_{nlJM}\ket{nlJM},
\end{equation}
with
\begin{eqnarray}\label{eq:2.13}
u_{nlJM} &=& \braket{nlJM}{\bm \rho} = \sum_{mq} C^{JM}_{lm1q} \braket{nlmq}{\bm \rho}\nonumber\\
 &=& \sum_{mq} C^{JM}_{lm1q} c_{nlmq}.
\end{eqnarray}
The last equality follows from the fact that the Clebsch-Gordan coefficient are real and from the orthonormality 
of the $g_{n-l,l}$ functions. The form of $\ket{\bm \rho}$ given by Eq.~\eqref{eq:2.12} contains the expansion 
of the vector density over the combined angular-momenta basis. It should be noted that the presence of the 
Clebsch-Gordan coefficients imposes that the values of $J$ and $M$ must satisfy the conditions \eqref{eq:2.10}. 
The Clebsch-Gordan coefficients impose also that the non-zero terms in the double sum of \eqref{eq:2.13} are 
such that $M = m+q$.

\begin{figure*}
\includegraphics[width = 0.9\textwidth]{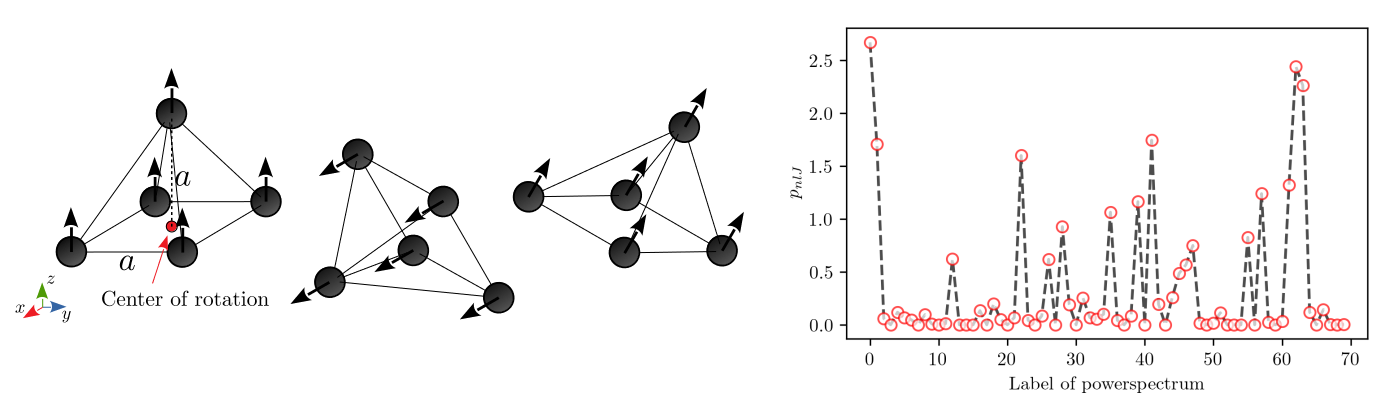}
\caption{\label{fig1}
(Color online) The invariant power spectrum (right-hand side panel) for a pyramid-shaped set of atoms, where a vector field 
is associated to each atom (left-hand side image). The power spectrum is evaluated with respect to the shown centre of 
rotation (red dot) and it is invariant against the simultaneous rotation of the system and the vectorial field around 
such centre. We chose the value $n_\text{max} = 6$ for this example. The number of independent power-spectrum 
components is given by $(n_\text{max}+1)(3n_\text{max}+2)/2 = 70$. The $x$-axis labels the power-spectrum 
components, while the $y$-axis shows their actual value.}
\end{figure*}

\subsection{Invariant power spectrum for a vector field}
In this section we are going to introduce an invariant power spectrum for the density given in Eq.~\eqref{eq:2.12}. 
Let us initially restrict our formulation to the case in which we ignore the atom at the origin of the local reference 
frame, namely we assume that the magnitude of the vector field is zero at the origin. One way to obtain the 
power spectrum, $p_{nlJ}$, is through the construction of the following inner product,
\begin{equation}\label{eq:2.14}
\braket{\bm \rho} = \sum_{nlJ} p_{nlJ}\:,
\end{equation}
which explicitly reads
\begin{equation}\label{eq:2.15}
p_{nlJ} = \sum_M \abs{u_{nlJM}}^2\:,
\end{equation}
where we have used the orthogonality of the $\ket{nlJM}$ basis. Since the vectors $\ket{nlJM}$ correspond
to the coupled angular momenta, they transform under system rotation, $\hat{R}$, as the spherical harmonics 
$Y^M_J$, namely
\begin{equation}\label{eq:2.16}
\hat{R}Y^M_J = \sum_{M'} D^J_{M'M}(\hat{R}) Y^{M'}_J\:,
\end{equation} 
where $D^J_{MM'}(\hat R )$ is the Wigner $D$-matrix associated to the rotation $\hat R$. It must be noted that, 
when one considers the original angular momentum basis, the rotation $\hat{R}$ appears as a simultaneous 
rotation of both the positions of the atoms and the vector field. By applying this rotation to the density in 
Eq.~\eqref{eq:2.12} we obtain
\begin{eqnarray}\label{eq:2.17}
\hat{R}\ket{\rho} &=& \sum_{nlJM} u_{nlJM}\hat{R}\ket{nlJM}=\nonumber\\
                  &=& \sum_{nlJM} u_{nlJM}\sum_{M'} D^J_{M'M}(\hat{R})\ket{nlJM'}=\nonumber\\
                  &=& \sum_{nlJM'} u_{nlJM'}'\ket{nlJM'}\:,
\end{eqnarray} 
from which we can infer the transformation rule for the expansion coefficients
\begin{equation}\label{eq:2.18}
\hat{R}: u_{nlJM} \rightarrow \sum_{M'} u_{nlJM'} D^J_{MM'}(\hat{R})\:.
\end{equation}
Therefore under rotation $\hat{R}$ the power spectrum, $p_{nlJ}=\sum_M u_{nlJM}^* u_{nlJM}$, transforms as
\begin{eqnarray}\label{eq:2.19}
\hat{R}: p_{nlJ} 	&\rightarrow&\sum_{M'M''}u^*_{nlJM'}u_{nlJM''}\underbrace{\sum_M\left(D^{J}_{MM'}\right)^*D^{J}_{MM''}}_{= \delta_{M'M''}}\nonumber\\
		&&= \sum_{M'} u_{nlJM'}^* u_{nlJM'}=p_{nlJ}\:,
\end{eqnarray}
where we have used the unitarity of the Wigner $D$-matrices [note that we have shorten the notation 
$D^{J}_{MM'}(\hat{R})$ into $D^{J}_{MM'}$]. This proves that the power spectrum obtained from 
Eq.~\eqref{eq:2.15} is rotationally invariant for simultaneous rotations of the atomic positions and the vector 
field. Such invariance is shown numerically in Fig.~\ref{fig1}, where the power spectra computed for 
different rotations are shown to perfectly overlap. \Review{In the Appendix, we will briefly discuss the generalized power spectrum 
connecting different radial channels}. Furthermore,
we will also extend our construction to the more general case of a tensorial density.

If we now take a localization function of the form 
\begin{equation}\label{eq:2.20}
h_a(\bm r-\bm r_i) = \delta(\bm r-\bm r_a)\:,
\end{equation}
namely a Dirac-delta function centered on the $a$-th atom, then the local vector density reads
\begin{equation}\label{eq:2.21}
\bm \rho(\bm r) = \sum_a w_a \delta(\bm r- \bm r_a)\bm v_a\:.
\end{equation}
In this case the expansion coefficients of Eq.~\eqref{eq:2.13} are readily evaluated by using Eq.~\eqref{eq:2.4} as
\begin{equation}\label{eq:2.22}
u_{nlJM} = \sum_a w_a g_{n-l,l} (r_a) \sum_{mq} C^{JM}_{lm1q}Y^{m*}_l(\hat{\bm r}_a)v_{a,q}\:.
\end{equation}\label{eq:2.23}
In what follows, we will use this expression to explicitly evaluate the power spectrum.

\Review{If we now consider a vector field having a vector-baring atom at the origin, it can be proven (see the 
Appendix) that the power spectrum is not generally invariant under rotations. We can interpret this rotational-symmetry 
breaking by noting that a vector field at the origin introduces an inner preferential direction for the local reference frame. 
One pragmatic solution to recover the invariance is to always rotate the system so that the vector field at the origin points 
along the $z$-axis. After such alignment we obtain a power spectrum, which is invariant under rotations around the $z$-axis.
Another possible solution, is to choose a suitable radial basis set so that all the non-invariant terms are automatically 
removed. As shown in the Appendix, we proved that the Spherical Bessel functions have this property. In the following, we 
will always consider power spectra with a central atom. As an example of its explicit evaluation, it is useful to obtain the 
complete expression for the $p_{n0J}$ power spectra. The component $u_{n0JM}$ is proportional to
\begin{equation}
u_{n0JM} \propto \delta_{J1}\sum_a w_a g_{n0}(r_a)v_{a,M}\:,
\end{equation}
where we used the equalities $q = M$ and $J=1$ enforced through the Clebsh-Gordan coefficients $C^{JM}_{0010}$, 
and we did not carry over the spherical harmonics values and the Clebsh-Gordan coefficients which, in this case, are 
unessential constants. The power spectrum is then proportional to 
\begin{eqnarray}\label{eq:2.27}
p_{n0J} &\propto& \delta_{J1} \sum_{ab} w_a w_b g_{n0} (r_a)g_{n0}(r_b) \sum_M v_{a,M} v_{b,M}^*\:,\nonumber\\
&\propto& \delta_{J1} \sum_{ab} w_a w_b g_{n0} (r_a)g_{n0}(r_b) \bm v_a \cdot \bm v_b\:.
\end{eqnarray}
It is interesting to note that, if the vector field is made of local spins, the $p_{n0J}$ power spectrum component will 
have a structure similar to that of an Heisenberg model, namely it depends on the inner product between the spins 
with distances-only dependent coefficients.}

Having derived a set of invariant power spectrum also for the case of an atom at the origin, we can now introduce 
the models used to test our formalism and the machine-learning scheme that implements the power spectrum.

\subsection{Training a machine-learning model}
The machine-learning model used here is a linear regression constructed over the power spectrum $\{p_{nlJ}\}$, 
following the same philosophy of SNAP \cite{SNAP}. Thus, let us assume to have a system of $N$ atoms, each 
one of them bearing a local spin. Let us also define the power spectrum vector, $\bm p$, as the one dimensional 
vector, whose entries are the $p_{nlJ}$ components describing a specific local neighborhood.  Specifically, the 
$i$-th power spectrum vector $\bm p ^{(i)}$ is the vector obtained by centring the local reference frame on the 
$i$-th atom, and then by evaluating the $i$-th power spectrum set $\{p_{nlJ}^{(i)}\}$ with respect to that frame. 
Thus, given our $N$-atoms system, we obtain $N$ power spectrum vectors, $\bm p^{(i)}$. 

Our main working hypothesis is that the energy, or any other quantity that we wish to represent, can be written 
as the sum of short-ranged contributions, $\varepsilon_i(\bm q^i)$, located on each atoms \cite{Bartok}, namely 
\begin{equation}
E = \sum_i \varepsilon_i(\bm q^i)\:,
\end{equation}
where $\bm q^i$ is a vector describing the local environment of the $i$-th atom. Then, following the same idea 
behind the SNAP \cite{SNAP}, we further assume that the power spectrum vectors $\bm p^{(i)}$ form a suitable 
set of descriptors to represent such decomposition, so that the local energies can be expressed as a linear 
combination of power spectrum vectors,
\begin{equation}
E \simeq \bm \theta \cdot\sum_i\bm p^{(i)}= \sum_{nlJ} \theta_{nlJ}\sum_i p_{nlJ}^{(i)}\:.
\end{equation}
Here $\{\theta_{nlJ}\}$ is an appropriate set of weights. The validity of these assumptions cannot be determined 
from the outset and must be tested on a case-by-case base. Within our formalism, the power spectrum vectors 
can be seen as the descriptors of a linear regression problem, where the target is the energy of the system. 

In what follows we will first calculate the energies of several atomic and spin configurations obtained by displacing 
the position of the atoms and the direction and magnitude of the magnetic vectors. Then, we will evaluate the 
power spectrum vectors for each atom and for each of the configurations considered. Finally, we will train a Ridge 
regression and optimize the weights vector, $\bm \theta$, to predict the total energy. This will allow us to investigate 
the descriptive power of our vectorial representation and of the full method proposed. In the next section 
we describe the different models investigated.

\begin{figure}
\includegraphics[width = 0.45\textwidth]{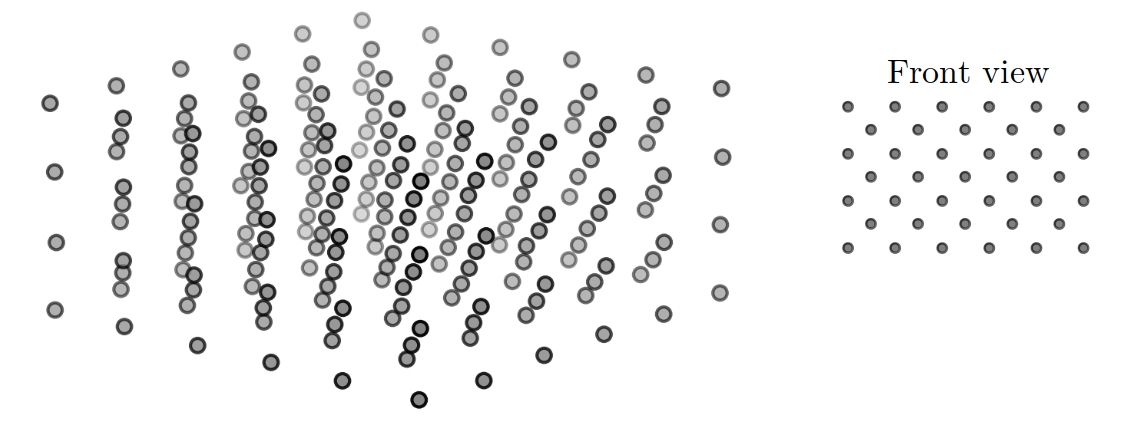}
\caption{\label{fig2}The physical system investigated in the present work: a tetrahedral cluster of {\it bcc} iron. 
The system is made of $7$ stacked square-shaped layers of $5$ or $6$ atoms per side, as shown in the insert. 
The total number of atoms is 219.}
\label{Fig2}
\end{figure}

\subsection{The physical system investigated}
In this work we consider a rectangular cluster of sites arranged over a {\it bcc} lattice, containing 219 atoms of the 
same species. The cell is rectangular with a six-atom wide square base and no periodic boundary conditions, as shown 
in Fig.~\ref{Fig2}. Each atom bares a local spin and can be displaced from the ideal high-symmetry {\it bcc} site. The 
training data, namely the atomic and spin configurations and their associated total energies, may come from a suitable 
total energy theory. This is usual some {\it ab initio} method such as spin-polarised density functional theory or a 
quantum-chemistry wave-function scheme. Since the generation of such dataset is rather time consuming, and our 
objective here is simply that of introducing our vector field representation, we use instead a range of analytical energy 
models.

In particular, we assume that the total energy is determined by the Hamiltonian,
\begin{equation}\label{eq:2.24}
H = H_\mathrm{H} + H_\mathrm{L}\:,
\end{equation}
where 
\begin{equation}
\begin{dcases}
H_\mathrm{H} = - \dfrac{1}{2}\sum_{\langle i,j\rangle} J_{ij}(r_{ij})\vb S_i\cdot\vb S_j\:,\\
H_\mathrm{L} = \sum_i\left( A S_i^2 + BS_i^4 + C S_i^6 \right)\:.
\end{dcases}
\end{equation}
Here, $H_\mathrm{H}$ describes an Heisenberg model, where the exchange parameter between the pair 
of atoms $\langle i,j\rangle$, baring spin $\vb S_i$ and $\vb S_j$, depends on the atoms distance $r_{ij}$. 
Note that the spin vectors are in units of $\hbar$, so that $S_i = M_i/g_e\mu_\mathrm{B}$, with $M_i$ being 
the $i$-th local magnetic moment and $\mu_\mathrm{B}$ the Bohr magneton. In particular, in this work we 
choose the following functional form~\cite{Ma2008} for $J_{ij}$ 
\begin{equation}
J_{ij}(r_{ij}) = J_{n}(1-\Delta r_{ij} /r_{n})^3, \qquad\text{with} \qquad \Delta r_{ij} = r_{ij} -r_n\:,
\end{equation} 
where the index $n$ indicates that the atoms $i$ and $j$ form a $n$-th neighbours pair. The distance $r_n$ 
is that between two $n$-th neighbour atoms in the undistorted {\it bcc} lattice (the $n$-th neighbour equilibrium 
distance). Similarly, the constants $J_n$ are the Heisenberg coupling elements between two $n$-th neighbours 
at equilibrium. It should be noted that $H_\mathrm{H}$ describes coupling between the position and the spin 
degrees of freedom by mean of the coupling constants, $J_{ij}(r_{ij})$. The Hamiltonian is then completed by 
a Landau-like term, $H_\mathrm{L}$, which describes the dependence of the energy on the longitudinal local 
magnetization (the magnitude of local spins)~\cite{PhysRevB.86.054416}, where $A, B$ and $C$ are constants
to be determined. 

In this work we set the various parameters to describe {\it bcc} iron~\cite{PhysRevB.86.054416}. Thus, the 
Heisenberg exchange interaction extends to second nearest neighbours with $J_1 = 22.52$~meV and 
$J_2 = 17.99$~meV, while the Landau parameters are chosen to be $A = -440.987$~{meV}, $B = 150.546$~{meV} 
and $C = 50.769$~{meV}. We will now proceed to show how our descriptors are able to capture the potential 
energy surface of the Hamiltonian of Eq.~(\ref{eq:2.24}).

\section{Numerical Simulations}

The Hamiltonian given in Eq.~\eqref{eq:2.24} consists in two qualitatively different terms, $H_\mathrm{H}$ and 
$H_\mathrm{L}$. The first describes transverse energy excitations and spin-lattice coupling, while the second 
accounts for longitudinal excitations. In order to investigate the descriptive power of the power spectrum and 
of our linear energy model over these two different types of excitation, we first consider only the Heisenberg 
term with fixed magnetic momenta lengths, $M_i= 2.2$ $\mu_\textrm{B}$, and later the full model. 

\subsection{Representing the Heisenberg Model with spin-lattice interaction}
The dataset has been built by displacing the atomic positions from the ideal equilibrium {\it bcc} structure 
and by choosing different orientations of the local magnetic moments. For the atomic positions we have 
chosen three sets presenting a different maximum displacement of 5\%, 10\% and 20\% of the lattice constant, 
respectively. The sampling of the displacements is uniform in space. In contrast, we have used two different 
strategies to define the spin structure. In the first one, we align the majority of the spins along the $z$-axis,
while the remaining magnetic moments point in a random direction. More specifically, out of the available 219 
magnetic moments, we randomly choose always more than 200 spins (the actual number is between 200 and 
219, and it also randomly selected), to be aligned along the $z$-axis, thus forming an almost-ferromagnetic 
structure (this training set is called the `ferromagnetic' one). The second strategy, instead, consists in assigning 
to all the magnetic moments a random orientation (`random' training set). Considering the three different choices 
for the maximum atomic displacement, and the two for the spins alignment, we have thus built a total of $6$ 
datasets, each made of $100$ configurations.

In order to test the predictions made by our fitted spin potentials, we build three further test-sets for each of 
the $6$ dataset explored. The first set consists of $219$ configurations; the $n$-th configuration having 
$n$ randomly-chosen spins aligned along the $z$-axis, while the remaining ones being randomly oriented in 
space. In this case the lattice is chosen to be pristine {\it bcc} (no atomic displacements) so to test independently 
the vectorial character of the potential. In contrast, the second and the third test-sets are designed to investigate 
also the atomic displacements. They consist of $50$ configurations each, and the atomic displacement has the same maximum magnitude of 
that of the dataset used to train the model. 
The spins configuration of the second test-set has 200 randomly-chosen spins aligned along the $z$-axis 
while the remaining ones are randomly oriented in space. The third test-set, instead, has all the spins randomly 
oriented in space. As such, these sets have been designed to test the predictions on a mostly-ferromagnetic 
environment and on a paramagnetic one, respectively. 
\begin{figure}[h]
\includegraphics[width = 0.48\textwidth]{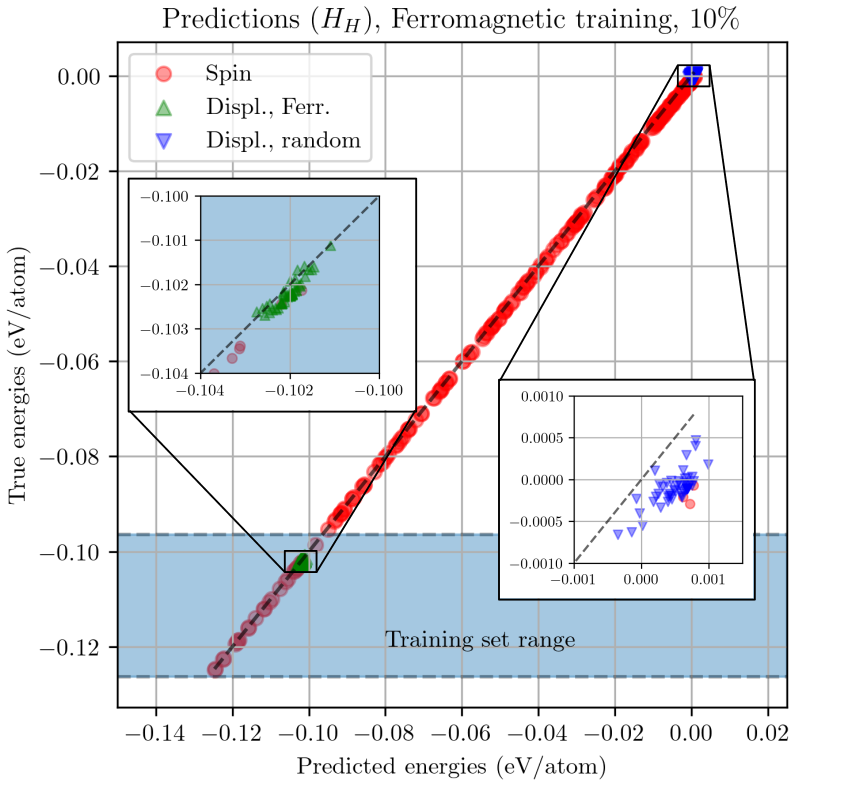}
\caption{\label{fig3} (Color online) Predicted against actual energies for a Ridge linear regression trained on the 
ferromagnetic dataset with $10$\% maximum atomic displacement. The actual energies are from the Heisenberg
model with spin-lattice coupling. The results are tested against three different test-sets. 
The red dots represent the configurations with undistorted {\it bcc} atomic positions and progressively $z$-aligned magnetic 
moments. The other dots represent different displacements of the atoms for near to ferromagnetic and paramagnetic
configurations, respectively. The figure demonstrates the good agreement of the potential also for configurations, 
which are energetically far from that used in the training (blue region). Zoom-in around different energy regions are
displayed in the inserts.}
\end{figure}

In Fig. \ref{fig3} we shown the results for the ferromagnetic-trained potential with a 10\% maximum atomic displacement,
for which we explicitly report the procedure and the results. The results on the other training sets are reported in the Supplementary Informations. The optimal potential parameters are found to be 
$n_\text{max}=4$, corresponding to 35 features only, and $r_\text{cut}= 1.4$ (lattice units), while the regularization 
constant of the Ridge-regression is $\alpha = \num{3.2e3}$.  
In the cross-validation procedure, we split the dataset in training- and test-sets five times 
with a 80-20 ratio with respect to the total dataset. We obtain an energy mean absolute error (MAE) of 
$(4.83\pm0.15)\times 10^{-5}$ eV/atom on the training set, and of $(6.8\pm0.7)\times 10^{-5}$ eV/atom on the test
one, which roughly corresponds to an error smaller than 0.1\%. When looking more specifically at the model 
predictions, the analysis on the first test-set (red points in Fig.~\ref{fig3}) returns us a MAE of $\num{5.6e-4}$~eV/atom
($\sim1$\%). Notably we find that the MAE of configurations with an energy above $-\num{0.01}$~eV/atom, namely 
those that are further away from the energy range of the training set, is $\num{8.2e-4}$~eV/atom. This means that the 
prediction of the model is still effective also in the portion of the configuration space far from that of the training. The 
MAEs on the second (green points in Fig \ref{fig3}) and third test-sets (blue points in the Figure) are respectively 
$\num{1.9e-4}$ eV/atom and $\num{6.0e-4}$ eV/atom. This suggests that the model can extrapolate rather well
across the configuration space.

The remaining trained models show us MAEs similar to that reported above, when the training is performed over the
ferromagnetic datasets. In contrast, \Review{when training on the three training sets denoted as  ``random'' spin 
configurations, we notice a significantly larger MAEs for large atomic displacements. Explicitly, the MAE is
$\num{1.3e-4}$~eV/atom for the random training set with 5\% maximum atomic displacement, but this already 
increases to $\num{2.5e-3}$~eV/atom for 10\% maximum displacement and reaches up to $\num{0.01}$~eV/atom 
for 20\% maximum displacement}. 
The failure for the largest maximum displacement can be attributed to the fact that the random dataset explores a much 
smaller portion of the energy landscape of the model. 
\Review{This is because the random configurations are all characterized by a small total magnetization and hence rather
similar energies.}
For all the other cases, the good agreement obtained between 
our model and the true potential energy surface, demonstrates that the Heisenberg model including spin-lattice coupling 
is accurately described by our potential, which is able to extrapolate the entire energy landscape. It is crucial to remark 
at this point that we did not introduce any prior knowledge 
of the functional dependence of the coupling constant on the pair-wise distance between the atoms, namely the model 
and the descriptors are able to autonomously interpolate the spin-phonon coupling. Moreover, given the modest size of the 
dataset, we found quite remarkable that simultaneous using a small datasets and a reduced number of features are able 
to reach the accuracies reported here.

Next we will consider the full Hamiltonian, including $H_\mathrm{L}$, describing both transverse and longitudinal spin 
excitations. 

\subsection{Heisenberg Model with longitudinal excitations}
The investigation of the model described by the complete Hamiltonian of Eq.~\eqref{eq:2.24} follows the same 
approach used for the analysis on the Heisenberg part. In this case we build a dataset corresponding only to one 
maximum displacement of the atomic positions, namely 10\% of the lattice constant. The spin configurations 
correspond to the ``ferromagnetic'' case described in the previous section. However, having to deal with longitudinal 
excitations as well, we also vary the magnetic moment's length. In particular, the magnetic moments aligned along 
the $z$-axis are chosen to be 2.25~$\mu_\textbf{B}$, while the randomly oriented ones have a length randomly 
chosen in the range 1.8-2.3~$\mu_\textbf{B}$.
\begin{figure}
\includegraphics[width = 0.48\textwidth]{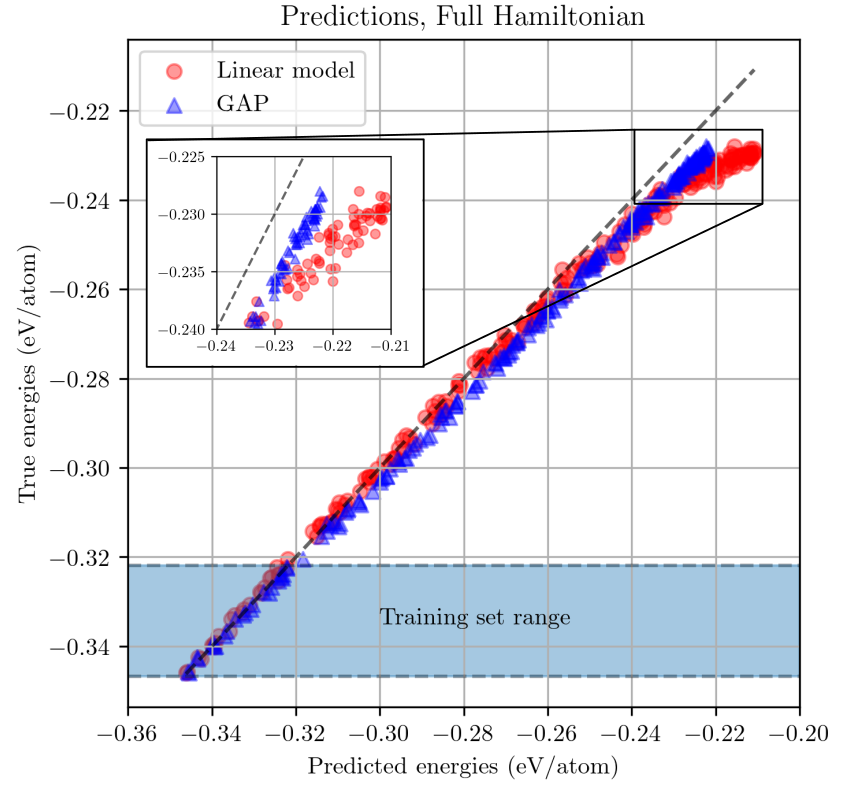}
\caption{\label{fig4}
(Color online) Predicted against actual energies for both a Ridge linear regression (red circles) and a 
GAP (blue triangles) constructed to predict the potential energy surface of the full Hamiltonian of Eq.~\eqref{eq:2.24} 
(Heisenberg model with spin-lattice coupling and longitudinal spin fluctuations). Results are presented for the test set. 
While the accuracy of the two models is similar for energies in the training set region, this becomes significantly different 
for the high-energy data points. In particular, the linear model accurately predicts energies up to about -0.26~eV/atom, 
but then significantly deviates from the parity line (dashed black line). In contrast, the GAP, while it appears to slightly
underestimate the actual energies, outperforms the linear model at extrapolating away from the training set region.}
\end{figure}

When testing the predictions, we build an additional test-set, corresponding to the first one presented in the previous section, 
namely containing an increasing number of aligning spins. In this case, the length of the $z$-aligned magnetic moments are 
again fixed to $2.25$ $\mu_\textbf{B}$, while the randomly oriented ones have a length in the range $1.9-2.3$ $\mu_\textbf{B}$.
Also the cross-validation procedure is similar to the one employed before with a five-time split of the dataset into training and 
test sets, with a 80-20 ratio. The parameters chosen are then $n_\text{max}= 4$, $r_\text{cut} = 1.4$ (lattice constant), 
$\alpha = \num{2e5}$. 

The MAEs obtained in this case are $(4.9\pm0.1)\times10^{-4}$~eV/atom and $(6.0\pm0.5)\times10^{-4}$~eV/atom, respectively 
for the training and test set. These values are about one order of magnitude larger than those obtained previously for the 
Heisenberg model with spin-lattice coupling. We can understand such accuracy loss by noticing that the descriptors are 
quadratic in the spin magnitude, as evident from Eqs. \eqref{eq:2.15}, \eqref{eq:2.22} and \eqref{eq:2.27}. Therefore, a 
linear machine-learning model, as employed here, will not be able to capture the energy contributions to the fourth and 
sixth power in the magnetization, which defines the longitudinal part of the Hamiltonian, $H_\textrm{L}$. In fact, it may be
surprising that the model still performs accurately even in this case. This is because we are exploring a region of the
potential energy surface relatively close to the minimum, where the energy contributions in $S^4$ and $S^6$ remain modest.  

In order to corroborate this hypothesis, we evaluate the model predictions on a test set containing progressively-aligned spins, 
for which we obtain a MAE of $\num{6.0 e-3}$ eV/atom. These results are shown as red dots in Fig.~\ref{fig4}, where
it is clear that the Ridge regression performs poorly as we progressively explore energy regions away from the training
range. Such behaviour must be associated to the limit of the machine-learning linear model constructed over our
description. In more detail, we find that the low-energy regions are still well described, with a MAE of $\num{7.1e-4}$~eV/atom 
for energies smaller than less than -0.32~eV/atom (in the training range). In contrast, the potential rapidly 
departs from the parity line at higher energies, where we compute a MAE of $\num{1.0e-2}$~eV/atom for data above 
-0.26~eV/atom.

We can improve on the error and go beyond the quadratic nature of our descriptors by combining the power-spectrum 
representation of the atomic and vector field with a non-linear machine-learning model. In particular, we consider here 
a Gaussian approximation potential (GAP)~\cite{gap}. GAP expresses the atomic energy of the $i$-th atom as
\begin{equation}
    \varepsilon_{i}=\sum_{t=0}^{N_\mathrm{train}}\theta_{t}S(\bm{p}^{i}, \bm{p}^{t})=\sum_{t}\theta_{t}e^{-{\frac{1}{2\sigma}(\bm{p}^{i}-\bm{p}^{t}})^{2}}
\end{equation}
where the sum is extended over all the atoms in the training set, and where $\bm{p}^{i}$ and $\bm{p}^{t}$ are
the power spectrum, respectively of the $i$-th atom and of the training set. The non linearity of the similarity 
kernel, $S$, allows us to describe energy contributions going beyond the quadratic order in the spin magnitude. 

The GAP predictions obtained over the test-set are shown in Fig.~\ref{fig4} as blue triangles. We notice that 
the MAE associated with the configurations having energy smaller than -0.32~eV/atom remains very close to 
that obtained with ridge regression. However, the total MAE decreases to $\num{4.5e-4}$~eV/atom and, most
importantly, the MAE for energies larger than -0.26~eV/atom is now reduced to $\num{5.8e-3}$~eV/atom, namely
is halved. In fact, the figure clearly shows that the non-linear GAP improves the ability of the model to extrapolate
away from the training set range. Since, the actual potential energy surface for a spin system, as the one obtainable
from density functional theory, is expected to include energy contributions going beyond a quadratic dependence
on the magnetization, we conclude that the best use of our representation will be in conjunction with non-linear 
machine-learning models.

\section{Conclusion}
In this work we have introduced a new invariant power spectrum representation for vectorial fields. 
After having presented an in-depth analysis of its rotational invariance and basic properties, we have 
designed a linear energy model, closely following the SNAP \cite{SNAP} approach. Such spin SNAP 
has then be put to the test against a {\it bcc} iron model described by an Hamiltonian containing 
spin-lattice coupling and both transverse and longitudinal energy excitations. Spin-lattice coupling 
is introduced by mean of a Heisenberg model with exchange parameters depending on the interatomic
distance, while the longitudinal spin excitations are described by a simple Landau term containing
even powers of the magnetization. 

We have then trained a first potential, linear in the power spectrum, for the situation where the longitudinal 
spin excitations are neglected. This was trained over a dataset obtained by displacing the atomic positions 
and the orientations of the atomic-magnetic moments, comprising a total of only 100 different configurations. 
Our results showed that the power spectrum is able to describe the entire energy surface, by accurately 
extrapolating far beyond the energy range  covered by the training set. This proves that a linear model 
using the power spectrum is sufficient to describe both the Heisenberg model and the spin-lattice coupling, 
already from a small dataset. Crucially, no prior information on the dependence of the exchange constants on 
the atomic position were used by the model. 

We have then repeated the exercise for the complete Hamiltonian, containing also the Landau term, by training 
over a dataset containing spins of different magnitude. Our results are highly accurate for configurations with 
energies within the range explored by our training set, but the model does not perform well in extrapolating. 
We have attributed this result to the inability of the power spectrum, combined with a linear machine-learning 
model, to describe energy contributions scaling beyond a quadratic dependence on the spin magnitude. Such 
shortcoming can be recovered by employing a non-linear model. Thus, we have investigated a Gaussian 
approximation potential and shown that extrapolation over a much-larger portion of the energy landscape is 
indeed possible. 

All in all, our analysis has shown that a power spectrum representation of the magnetization field can be
used, together with non-linear machine-learning models, as an efficient descriptor of spin potential
energy surfaces. This can now be used in conjunction with training sets obtained from accurate electronic
structure theory to predict finite temperature properties of magnets.

\begin{acknowledgments}
This work has been supported by the Irish Research Council Advanced Laureate Award (IRCLA/2019/127),
and by the Irish Research Council postgraduate program (MC). We acknowledge the DJEI/DES/SFI/HEA Irish 
Centre for High-End Computing (ICHEC) and Trinity Centre for High Performance Computing (TCHPC) for the 
provision of computational resources. 
\end{acknowledgments}

\appendix
\Review{
\section{Atom-centered power spectrum}\label{App0}
Let us prove that, if the vector field has an atom at the origin, the power spectrum of \eqref{eq:2.15} will not be 
generally rotationally invariant. For simplicity we can consider first the trivial case in which our system is made 
of just a single atom, e.g. there are no other atoms inside the cut-off radius. The component $u_{n100}$ of the 
coefficients of Eq.~\eqref{eq:2.22} reads
\begin{eqnarray}\label{eq:2.23A}
u_{n100} &=& w_0 R_{n1}(0)\sum_{mq}Y^{m*}_1(\bm 0)C^{00}_{1m1q} v_q\:,\nonumber\\
&\propto& R_{n1}(0)\sum_{q}C^{00}_{101q}v_q\:,\nonumber\\
&\propto& R_{n1}(0)v_0\:,
\end{eqnarray}
where in the first step we have used the fact that the spherical harmonics along the $z$-axis vanish unless $m=0$, and 
in the second step we have considered the equality $q=M-m$ implemented through the Clebsch-Gordan coefficients
(for simplicity we do not carry over unessential constant). The power spectrum component for this case then is simply
\begin{equation}\label{eq:2.24A}
p_{n10} \propto \abs{R_{n1}(0) v_0}^2\:.
\end{equation}
Equation~\eqref{eq:2.24A} establishes that the $n10$ component of the power spectrum is proportional to the 
magnitude of the $z$ component of the vector field at the origin. Crucially, for a general rotation of the reference 
frame, the $z$ component becomes mixed with the other components, so that it changes its value. We then 
deduce that the power spectrum is not rotational invariant, if the vector field does not vanish at the origin of the 
local reference frame. It is worth stressing that this proof holds, since the origin is a fixed point for the rotation. 
If the center or rotation is not the origin, then also the argument of the spherical harmonics will rotate, making the 
$m\neq 0$ terms relevant too. This proof can be generalized also to the case in which there are other atoms 
within the cutoff radius. 
As pointed out in the main text, we can however recover a cylindrical symmetry by simply rotate the system so that the vector fields in the origin points in the same direction as the $z$-axis. Another possible solution stems from 
the presence of the radial-basis set in Eq.~\eqref{eq:2.24A} which, if carefully chosen, could remove the symmetry breaking terms of the power spectrum. Spherical Bessel functions are suitable for this 
purpose, as we will show now.
We first notice that for $l\neq 0$ the function vanishes at the origin, namely $g_{n-l,l}(0)= 0$. This is a consequence 
of the fact that the basis set is defined in terms of the spherical-Bessel functions $j_l(x)$, which also vanish at the 
origin for non-zero $l$. Thus, for $l\neq 0$ all the contributions arising from an atom at the origin are removed, 
ensuring the rotational invariance. We are then left to prove that the power spectrum is invariant also for $l=0$. 
This is easily done by noticing that the spherical harmonic $Y^0_0$ is just a constant, with no angular dependence, completing the demonstration.
}
\section{Power spectrum connecting different radial channels}\label{AppA}
Following the same approach used in reference \cite{Bartok}, we can generalize the expression for the 
power spectrum \Review{so that different radial channels are coupled.}
A generalized expression for the power spectrum may read
\Review{
\begin{equation}
p_{nn'lJ} = \sum_M u_{nlJM}^* u_{n'lJM}\:.
\end{equation}
}
The rotational invariance is still ensured by the fact that the transformation rules for the expansion 
coefficients $u_{nlJM}$ involve only the Wigner D-matrix belonging to the angular momentum $J$ 
[see Eq.~\eqref{eq:2.17}]. \Review{However, it is apparent that the number of components defining the descriptors are increased with this coupling choice. Note that we must ensure the same $l$ for the two factors in the sum above, so that the resulting quantity is real: the only complex part are found in the angular dependent terms being the radial functions real.}

\section{Generalization to a tensorial density}
In this appendix we generalize our formalism from a vectorial field to a tensorial one. In this case, the local density reads
\begin{equation}
\bm \rho(\vb r) = \sum_a w_a h_a(\vb r-\vb r_a) \bm T^{n_a}_a\:,
\end{equation}
where $\bm T^{n_a}_a$ is a $n_a$-rank tensor associated to the $a$-th atom. Note that tensors associated with different 
atoms can have different ranks \Review{as, for example, when dealing simultaneously with a scalar field and a vector field}. We can then use the spherical decomposition of the tensors and write $\bm T^{n_a}_a$ in 
Dirac notation as
\begin{equation}
\ket{\bm T^{n_a}_a} = \sum_\lambda^{n_a}\sum_{\mu= -\lambda}^\lambda (T_a)^\mu_\lambda\ket{\lambda\mu}\:,
\end{equation}
where $(T_a)^\mu_\lambda$ is the spherical component of the tensor $\bm T^{n_a}_a$ relative to the spherical basis 
$\ket{\lambda\mu}$. This is analogous to $v_{a,q}$ and $\ket{q}$ for the case of a vectorial field. The components 
$(T_a)^\mu_\lambda$ transform as the the spherical harmonic $Y^\mu_\lambda$ under rotation. We can then write 
the density as
\begin{equation}
\ket{\bm \rho} = \sum_{n=0}^{n_\text{max}}\sum_{l=0}^n\sum_{m=-l}^l\sum_{\lambda=0}^{\Lambda}\sum_{\mu= -\lambda}^\lambda c_{nlm\lambda\mu}\ket{nlm\lambda\mu}\:,
\end{equation}
where the expansion has been truncated at $n_\text{max}$, and with $\Lambda = \max_a(n_a)$ being the highest rank 
of the tensors in the tensorial field \Review{additional zero coefficients can be introduced to have an homogeneous representation in the highest-tensorial rank}. The expansion coefficients are then obtained by projection as
\begin{equation}
c_{nlm\lambda\mu} = \sum_a w_a (T_a)^\mu_\lambda \int \dd r\dd \hat{\bm r}\: r^2 R_{nl}(r)Y^{m*}_l(\hat{\bm r}) h_a (\bm r-\bm r_a)\:,
\end{equation}
with $(T_a)^\mu_\lambda = 0$ if $\lambda > n_a$. Following the same procedure outlined previously for the case of a vectorial 
field, we can express the density over the coupled basis as
\begin{equation}
\ket{\bm \rho} = \sum_{nl\lambda JM} u_{nl\lambda JM}\ket{nl\lambda JM}\:,
\end{equation}
by mean of the coupling scheme,
\begin{equation}
\ket{nl\lambda J M} = \sum_{m=l}^l \sum_{\mu = -\lambda}^\lambda C^{JM}_{lm\lambda\mu}\ket{nlm\lambda \mu}\:,
\end{equation}
with $\abs{l-\lambda}\le J \le l+\lambda$. The coupled-basis coefficients are given in terms of the uncoupled ones as
\begin{equation}
u_{nl\lambda JM} = \sum_{m\mu} C^{JM}_{lm\lambda\mu} c_{nlm\lambda\mu}\:.
\end{equation}

Finally, the power spectrum is again given by squaring the coefficients
\begin{equation}
p_{nl\lambda J} = \sum_M \abs{u_{nl\lambda J M}}^2,
\end{equation}
and it is invariant under simultaneous rotations of the frame of reference and the tensorial field. \Review{A further 
generalization to multi-channel coupling can be obtained by using the same argument
presented in Appendix~\ref{AppA}}.

\end{document}